# Probing pair-breaking mechanisms in proximity-induced hybrid superconducting interfaces

S. Mohapatra, S. Mathimalar, S. Chaudhary, K. V. Raman

*Centre for Interdisciplinary Sciences,*
*Tata Institute of Fundamental Research, Hyderabad 500107, India*

Corresponding author: kvraman@tifrh.res.in

**Abstract**

**Understanding depairing effects in a hybrid-superconducting interface utilizing high spin-orbit materials such as topological insulators or 1D semiconducting nanowires is becoming an important research topic in the study of proximity-induced superconductivity. Experimentally, proximity-induced superconductivity is found to suppress at much lower magnetic fields compared to the superconducting layer without a good understanding of its cause. Here, we provide a phenomenological tool to characterize different pair-breaking mechanisms, the ones that break or preserve time reversal symmetry, and show how they affect the differential tunneling conductance response. Importantly, we probe the properties of the SC layer at the hybrid interface and observe conductance peak pinning at zero bias in a larger field range with eventual signs of weak peak splitting. Further, the effect of varying the spin-orbit scattering and the Lande *g-factor* in tuning the conductance peaks show interesting trends.**

Recent studies of proximity induced superconductivity (PIS) at the interface between a s-wave superconductor (SC) and a topological material (TM), such as 3-Dimensional (3-D) topological insulators (TI)[1], 2-D TI layers[2] or 1-D high spin-orbit (SO) semiconducting nanowires[3] is laying a new scientific foundation bed for the exploration of conceptually rich condensed matter phenomena, especially in the quest for topological superconductivity. Here, the PIS is known to lead to three important consequences. Firstly, it leads to suppression of superconductivity in the SC layer, giving rise to a reduced superconducting gap ($\Delta_r$) near the interface compared to the bulk of SC ($\Delta_o$)[4] (Fig. 1a). Secondly, and more interestingly, it induces superconducting correlations in the adjacent layer, with the pair potential dropping discontinuously across the interface[5]. This induced condensate pair potential, $\Delta_i$, then decay with a characteristic length determined by the strength of decoherence or depairing properties of the induced layer. Such a behavior at such hybrid interfaces using TMs has been experimentally observed using angle resolved photoemission spectroscopy (ARPES) and point contact spectroscopy[6,7]. Lastly, the microscopic nature of the PIS condensate pairs may differ from the singlet s-wave pairing observed in the bulk of SC films. For example, in the case of ferromagnetic films in proximity to the SC layer, formation of triplet pairing is reported that is already contributing to the field of superconducting spintronics[8]. However, in the case of TMs, currently, very little is understood about the superconducting nature of these PIS states. In this regard, Blonder-Tinkham-Klapwijk (BTK)[9] formalism that considers anisotropic pairing potential with different orbital symmetries has been widely used to fit the differential conductance measurements using tunneling and point contact spectroscopy[6,7,10]. Here, the BTK model uses a phenomenological Dyne's parameter[11], $\gamma$, to incorporate depairing effects caused by finite lifetime of the quasiparticle excitations. However, it is expected that in TMs, these condensates can experience other forms of depairing effects: one that preserves time-reversal symmetry (TRS) such as spin-orbit (SO) scattering which contributes to depairing only in presence of applied magnetic field and one that breaks TRS in zero field such as orbital depairing and magnetic impurity scattering. Thus, the BTK model does not provide a complete microscopic picture in such systems where such depairing effects may contribute differently to suppression of superconductivity.

In the PIS study with TMs, experimental visualization of topological superconductivity to realize Majorana modes as zero-energy quasiparticle excitations in differential tunneling conductance spectroscopy has been widely attempted[10,12,13]. This is primarily predicted in materials where the bulk spin-orbit strongly mixes the two spin sub-bands giving rise to a topological band with a superconducting gap[1,3]. This gap, in an increasing applied magnetic field, is expected to close and cause band inversion, subsequently leading to a topological gap opening which supports zero energy Majorana modes[12]. In many experimental results, a 'rigid' zero bias peak is often observed, associated to the Majorana modes, that does not get affected over a large range of external stimulus such as applied magnetic field or gate voltage[10,12]. Such a response is shown to be very different from the other mechanisms that also give rise to a zero energy quasi-particle excitations such as Kondo peak, Andreev bound states, reflectionless tunneling and weak antilocalization. Thus, many of these reports have made a major advancement in our understanding of topological superconductivity; however, they have also left open some interesting questions regarding the device's conductance response to the above external stimulus. For example, in some studies[12], with the increase in magnetic field, the ZBCP is observed before gap closing which then disappear without any signs of topological gap opening. Instead, the ZBCP is found to weakly split into two peaks before the PIS is completely suppressed. Also, the PIS states are observed to disappear at small fields. For example, in the bilayer study of $NbSe_2/Bi_2Se_3$ films, the induced superconductivity disappears at just 0.03T compared to a 4T field needed to destroy superconductivity in the $NbSe_2$ layer[6,7]. Therefore, probing the properties of such hybrid SC/TM interfaces may allow a better understanding of PIS that is stable at higher magnetic fields.

In this article, we focus our attention on two aspects of the hybrid SC/TM interfaces. Firstly, how does the property of the SC film, driving the PIS state, gets affected due to the proximity effect. Secondly, can the response of the induced quasiparticle excitations be explained using a model better than the BTK formalism that can capture the microscopic origin of the various pair-breaking mechanisms responsible for the stronger suppression of induced superconductivity. In this regard, a systematic experimental study to characterize such hybrid interfaces is missing. We therefore focus on a device geometry as shown in Figure 1a in a strong tunnel barrier regime, neglecting any contributions due to Andreev bound states, Kondo or weak-anti-localization to the differential conductance. To explore such devices, we adopt a theoretical framework in modeling the effective quasiparticle density of states by using the Green's function approach developed first by Maki[14,15] and later generalized to model s-wave SC films with thickness lower than the penetration depth. This is done by considering separate Green's function for spin-up and spin-down superconducting electrons in an external Zeeman field, with the inclusion of impurities (magnetic and spin-orbit) as scattering centers within the SC region[16,17,18,19]. Ref. [19] provides a detailed analysis of the Green function formalism, giving us the following form of the quasiparticle density of states ($\rho$) for the spin-down ($\downarrow$) and spin-up ($\uparrow$) states:

$$\rho_{\downarrow\uparrow}(E) = \frac{\rho_0}{2} sign(E) Re\left\{\frac{u_\pm}{(u_\pm^2 - 1)^{1/2}}\right\} \qquad - Eq. 1$$

where $\rho_0$ is the normal density of states, E is the energy with respect to Fermi level, and $u_\pm$ are the complex energy functions for the spin down(+) and spin-up states(−) represented as:

$$u_\pm = \frac{(E - i\frac{E}{|E|}\Upsilon) \mp e_z}{\Delta} + \frac{\zeta u_\pm}{\Delta(1-u_\pm^2)^{1/2}} \mp b_{so}\left\{\frac{u_+ - u_-}{\Delta(1-u_\mp^2)^{1/2}}\right\} \pm d_{sf}\left\{\frac{u_+ + u_-}{\Delta(1-u_\mp^2)^{1/2}}\right\} \quad - Eq. 2$$

Here, $\Delta$ is the superconducting energy gap, $\zeta$, $b_{so}$, $d_{sf}$ are the parameters related (inversely) to the orbital depairing, spin-orbit and spin-flip scattering lifetimes[16,17], respectively, and the Zeeman energy ($2E_z$) is given by $g_{eff}\mu_B H$ with H as the applied magnetic field in the plane of the SC film and $g_{eff}$ as the effective Lande *g-factor*. It is important to note that density of states in equation (1) reduces to the theoretical form of BCS density of states in the absence of the above four depairing effects and Zeeman field. Numerical methods to solve the above coupled complex equations is known to be non-trivial due to the singularities at $|u_\pm|=1$ [20] and due to the inability to generate an analytical closed-form solution[21]. Limited attempts have been made earlier to solve the above set of coupled non-linear complex functions with only the orbital depairing and spin-orbit terms by using the Fermi liquid approaches under the dirty limit[20,22]. We extend similar approach to linearize the full form of the above coupled complex energy functions in *equation 2* to successfully determine the physical solution of $u_\pm$ for real values of the density of states. This leads to a set of four linearized equations given by:

$$(E - i\frac{E}{|E|}\Upsilon)y_2 + y_1 + y_4 e_z^2 - \frac{\zeta}{\pi}(y_1 y_2 - y_3 y_4 e_z^2) - \frac{2d_{sf}}{\pi}(y_1 y_2 + y_3 y_4 e_z^2) = 0$$

$$(E - i\frac{E}{|E|}\Upsilon)y_4 + y_2 - y_3 + \frac{\zeta}{\pi}(y_2 y_3 - y_1 y_4) - \frac{2b_{so}}{\pi}(y_1 y_4 + y_2 y_3) = 0$$

$$y_1^2 - y_2^2 \Delta^2 + y_3^2 e_z^2 - y_4^2 e_z^2 \Delta^2 + \pi^2 = 0 \qquad - Eq. 3$$

$$y_1 y_3 + y_2 y_4 \Delta^2 = 0,$$

where $y_i$; $i \in [1,4]$, represents the four complex variable related to the complex energy functions by the following expression:

$$y_1 \pm e_z y_3 = -\pi u_\pm (1 - u_\pm^2)^{-1/2}$$
$$\Delta(y_2 \mp e_z y_4) = \pi(1 - u_\pm^2)^{-1/2}, \qquad - Eq. 4$$

This procedure of arriving at solution to *Eq. 2* results in 8 sets of complex solutions, seven of which are discarded due to triviality (e.g. 0 or negative value of density of states from *Eq. 1*). The differential tunneling conductance for the device structure in Figure *1a* is then determined at any finite temperature using the above solution of the quasi-particle density of states. Figure 2 shows such a conductance map at 30mK with increasing strength of the four interfacial pair breaking mechanisms in zero and applied magnetic field; each of which show a characteristic response. Here, the SC film thickness is considered to be smaller than the penetration depth, leading to Zeeman splitting of the quasi-particle states in applied magnetic field[23]. Firstly, we observe that reducing the quasiparticle lifetime (i.e. increasing $\zeta$), leads to emergence of states within the gap (dark blue, representing a hard gap, fading to lighter shades at zero voltage bias), which may be associated to soft gap or gapless superconductivity. Furthermore, they lead to broadening of the conductance peak. Next, the orbital depairing is observed to only broaden the conductance peaks, while the spin-scattering causes both- broadening of the peaks and spins flipping that strongly suppress superconductivity with the emergence of weak gapless superconductivity. In contrast, the device conductance response to an increasing spin-orbit scattering, which preserves TRS, shows an interesting trend. Here, no effect is observed on the device conductance in zero field. However, in the presence of a magnetic field, it contributes to a weak peak broadening while maintaining a hard gap. Further, due to stronger spin-mixing, it counteracts the effect of Zeeman field on the quasiparticle states, making it harder to achieve the spin-splitting (in Figure 2h, the red feature shows a negative slope). Understanding the contributions from each depairing term hence provides a comprehensive and powerful methodology to fit experimental data to extract important microscopic information about the superconducting properties of the interface layers.

In the context of an SC/TM hybrid interface, the effect of the spin-orbit interaction and Zeeman field on the properties of the quasiparticle excitations in the SC and TM layers may require careful analysis. In the case of TM layer, contributions from a bulk spin-orbit coupling or Rashba spin-orbit coupling is expected to give rise to a topological electronic band structure, confirmed from ARPES studies[6]. However, due to stronger suppression of induced superconductivity at low applied magnetic field, there is still no clear evidence of topological gap opening in transport measurements (for SC/TM/I/NM device geometry as shown in Figure 1b). Our model, as a fitting routine, can therefore help resolve the dominant depairing contributions affecting these PIS state, a study that is yet to be explored. Regarding the SC layer, very little is experimentally probed (using TM/SC/I/NM device geometry in Figure 1b) about its properties due to the proximity with the TM layer. Firstly, since the superconducting condensates leak into the TM layer, we expect depairing effects, attributed mostly to spin-orbit scattering, to suppress superconductivity. Further, in the case of 1-D semiconducting nanowires as TM layers, the SC condensates may experience a larger value of $g_{eff}$[24,25] that offer significant Zeeman interaction energy at relatively low magnetic fields that, interestingly, are not strong enough to suppress superconductivity in the SC layer. Additionally, proximity effect and/or interface structural disorder may also substantially enhance the value of penetration depth (more than 200 nm)[26,27] causing Zeeman-splitting of the quasiparticle density of states in thicker SC films.

In the subsequent study, we devote our attention to investigate the superconducting properties of the SC layer by simulating the differential conductance response (see Fig. 3) of the NM/I/SC/TM device in *Figure 1b* with increasing in-plane Zeeman field for two different depairing terms, $\gamma$ and $b_{so}$ (see Fig. S1 for $\zeta$, $d_{sf}$). We assume a $g_{eff}$ of 10. Although, as discussed above, the dependence of $\Delta$ on the magnetic field is more complex, we here assume $\Delta$ to decrease by a second order transition as $\Delta(H) = \Delta(0)(1 - (H/H_c)^2)^{1/2}$, where $H_c$ is the critical magnetic field and $\Delta(0)$ is the zero field SC gap. Maki and Tsuneto[14] had shown that in the absence of depairing contributions, below the bulk critical temperature, the transition from a superconducting to normal state (due to magnetic field) moves from a second order to first order at a particular transition temperature. However, the presence of strong depairing mechanisms can significantly reduce this transition temperature or perhaps also completely suppress them leading to only second-order transitions at low temperatures[21]. Similar responses have been observed experimentally[10], but require further experimental verifications.

Figure 3a & 3c shows that with increasing Zeeman field, the device's zero bias conductance gradually shift from a hard gap to a soft gap and finally to a peak prevailing over a finite field range (~100mT, 0.7T to 0.8T in Fig. 3a, ~0.8T to 0.9T in Fig. 3c) until the ZBCP eventually splits. Further, the energy position of higher energy conductance peak is observed to increase initially and later fall down with applied magnetic field until the critical field. We further observe, as shown in figure 3b, that increasing $\gamma$ leads to disappearance of 'hard gap' (even in zero field) and, due to the difficulty in resolving the ZBCP splitting, makes the ZBCP exist over a larger field range (~200mT, ~ 0.7T to 0.9T). An increase in SO scattering also enhances the field range (~200mT, ~ 0.9T to 1.0T) for the observation of ZBCP as it makes Zeeman splitting harder to achieve. However it has one distinction compared to the other depairing terms; it retains the 'hard gap' at lower fields (Fig. 3c & d). Some of these characteristic responses are also seen in experiments, however with the interest in probing the response of induced superconductivity. Therefore, our analysis calls for careful scrutiny in interpretation of experimental studies using differential tunneling conductance measurements. We would like to emphasize that for Zeeman energy (primarily due to large $g_{eff}$) approaching Δ, our model assumes the ground state of the interface state to have a homogeneous order parameter. Transition to in-homogeneous superconducting state, such as Fulde, Ferrell, Larkin and Ovchinnikov (FFLO) states[28] is not considered since these states are very sensitive to disorder and are expected to occur only in very clean systems[21].

Next, we probe the ZBCP state in applied field and simulate the response of the SC layer in the hybrid SC/TM interface to the modulations in $b_{so}$. Experimentally, gate voltage modulation may also indirectly affect the strength of the SO scattering potential due to the adjacent TM layer (see Figure 1b). Figure 4a shows the variation in the differential conductance contour taken at an applied field of 0.85T. Interestingly, we observe the ZBCP (and its weak splitting) to gradually disappear, with the reappearance of soft gap and eventually a hard gap. Additionally, recent work in 1-D nanowire systems has suggested a strong anisotropy of $g_{eff}$ depending on the direction of the applied magnetic field with respect to the nanowire axis[29]. Furthermore, it is proposed that the gate voltage modulation can also affect the value of $g_{eff}$. Hence, in figure 4b, we model such a scenario by varying the value of $g_{eff}$ from 2 (lower limit as in bulk SC) to 10 that may be caused by leakage of SC condensates in the nanowire. Interestingly, we observe a similar trend of disappearance of the ZBCP with reducing $g_{eff}$, leading to angular or gate-modulation dependence, primarily due to lower Zeeman energy that reduces the spin splitting of the quasiparticle states. Thus our model, provide a good fitting tool with a focus on understanding the microscopy origin of depairing effects that limit superconductivity in the SC and TM layers of the hybrid superconducting interface. We expect such a fitting routine to help understand a number of exotic interface studies including the interface proximity study using ferromagnetic insulator where magnetic impurity governed depairing effects (parameter $d_{sf}$) may play a dominant role.

In conclusion, our work highlights the distinct characteristic response of different depairing interactions to differential conductance measurements in presence of external stimulus. Further, it is known that SO mechanisms can lead to exotic behaviors, such as the possibility of co-existence of superconductivity and magnetism at these interfaces. Our model can also study such a phenomena, i.e. co-existence of proximity induced superconductivity and ferromagnetism, by considering an internal exchange field acting on the SC condensates[30] (see Fig. S4). Future developments in the presented model approach to incorporate these depairing effects, within a unified formalism covering barrier-less to weak-barrier conductance for point contact spectroscopy measurements, will advance our understanding of PIS in these new classes of hybrid interfaces.

## Acknowledgements

KVR acknowledges TIFR, SERB Ramanujan and Early Career research grant for funding. SM acknowledges SERB NPDF fellowship.

# Figures

**Figure 1.**

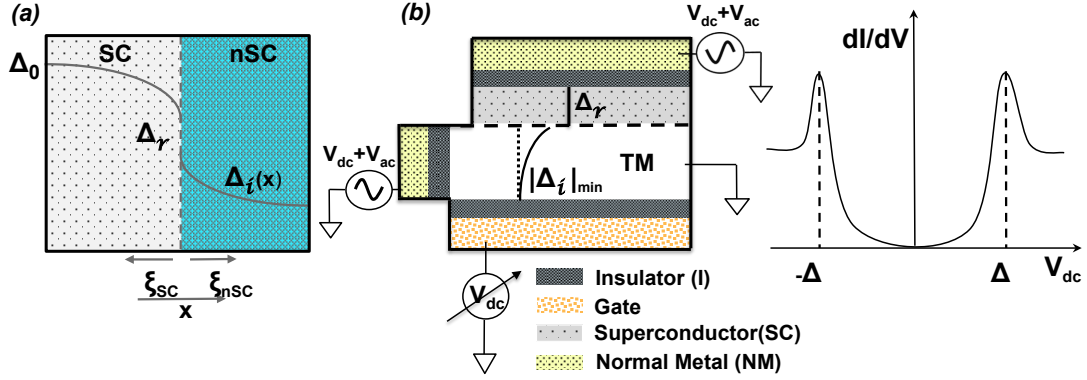

**Figure 1: Schematic showing the spatial variation of superconducting pair potential at the hybrid SC/TM interface** *(a)* SC order parameter ($\Delta$) reduces from its bulk value, $\Delta_0$, to a reduced value, $\Delta_r$, at the hybrid interface, followed by a discontinuous jump to $\Delta_i$ and decay into the TM layer. Such interface effects happen in the length scales of the coherence length ($\xi$). *(b)* Device configuration probing the quasiparticle response in the TM and the SC layer separately by measuring differential conductance through the two circuitry: (i) NM/I/SC/TM, and (ii) NM/I/TM tunnel junctions. Here, the SC layer thickness is smaller than $\xi$. Tunneling conductance of the devices show a conductance peak at $\pm|\Delta|$ associated either to $\Delta_r$ or $\Delta_i$.

**Figure 2.**

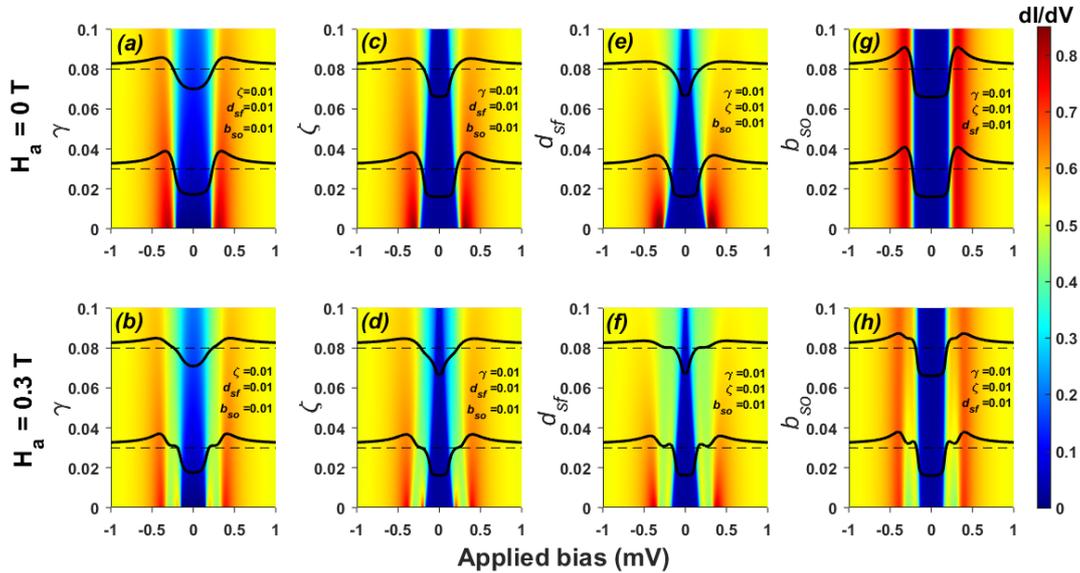

**Figure 2:** Simulated normalized differential conductance (dI/dV, shown as colorplots in arbitrary units) vs applied bias voltage of an TM/SC/I/NM device structure for the different pair breaking parameters, *(a)* and *(b)* for Dyne's parameter, $\gamma$, *(c)* and *(d)* for orbital depairing, $\zeta$, *(e)* and *(f)* for spin-scattering, $d_{sf}$, *(g)* and *(h)* for spin-orbit scattering, $b_{so}$, in zero magnetic field (top plots) and an applied magnetic field of 300 mT (bottom plots). Superimposed dI/dV line plots (solid line) represent the differential conductance at the parameter values corresponding to the horizontal line cuts (dashed line). For each subfigure, values of other depairing parameters are kept at 0.01, $\Delta(0) = 0.3$ meV, $g_{eff} = 10$ and $T = 30$ mK.

**Figure 3.**

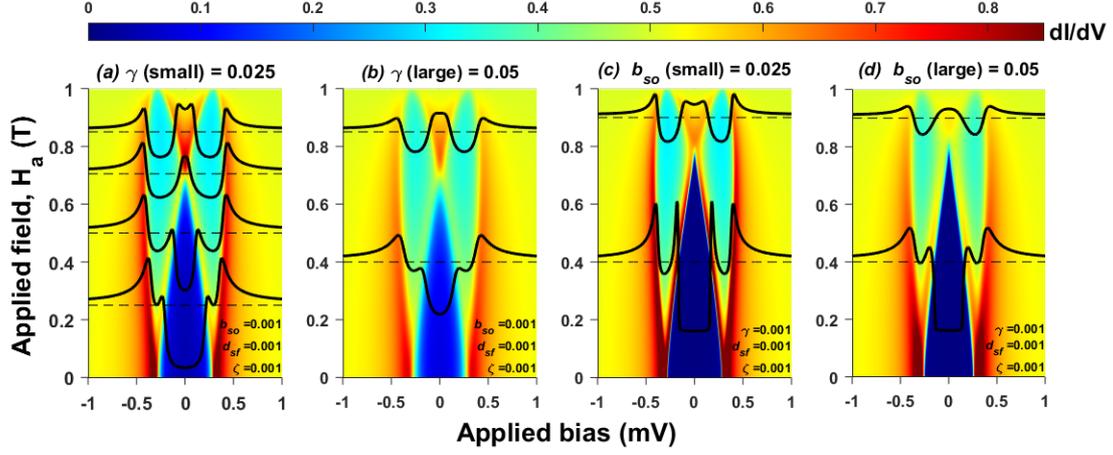

Figure 3: Simulated normalized differential conductance (dI/dV, shown as colorplots in arbitrary units) vs applied bias voltage of an TM/SC/I/NM device for varying applied magnetic field with $g_{eff}$ =10, $\Delta(0) = 0.3$meV, $H_c$ = 1T & $T$ = 30mK, plotted for increasing depairing contributions of Dyne's parameter (*a* & *b*) and spin-orbit scattering (*c* & *d*). With increasing field, hard gap (dark blue) moves to a soft gap (light blue) and eventually to a ZBCP (red). In *(a)* and *(c)* splitting of ZBCP is visible at higher fields, which cannot be resolved with the increase in depairing contributions (*b* & *d*). Superimposed dI/dV line plots (solid line) represent the dI/dV at the field values corresponding to the horizontal line cuts (dashed line).

**Figure 4.**

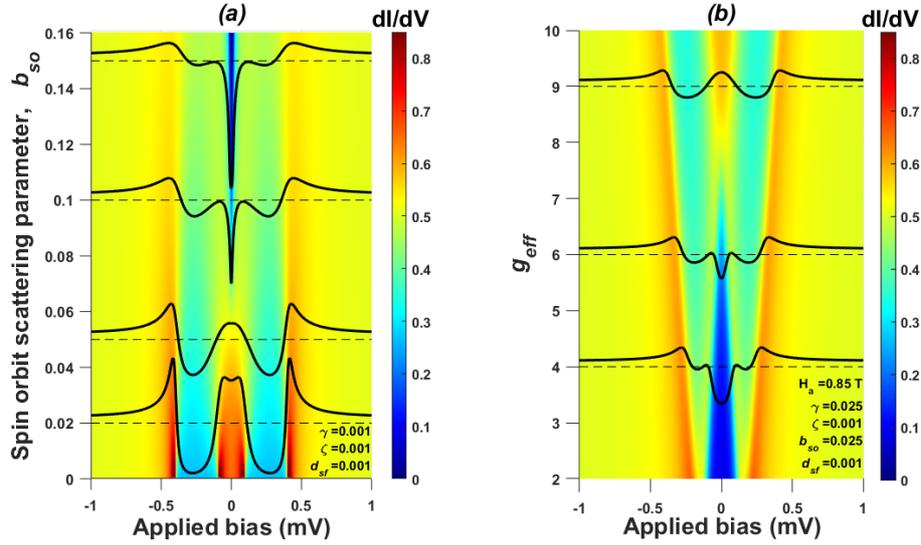

**Figure 4:** Simulated normalized differential conductance (dI/dV, shown as colorplots in arbitrary units) vs applied bias voltage of an TM/SC/I/NM device with *(a)* increasing strength of spin-orbit scattering, $b_{so}$, with $g_{eff}$ =10, $H$=0.85T, $\Delta(0) = 0.3$meV, $H_c$=1T & $T$=30mK and *(b)* increasing value of effective Lande g-factor, $g_{eff}$; with $H$=0.85T, $\Delta(0) = 0.3$meV, $H_c$=1T & $T$=30mK. In *(a)*, with increasing $b_{so}$, ZBCP disappears and a hard gap appears at higher $b_{so}$ (see Figure S2). In *(b)*, ZBCP appears with increase in $g_{eff}$ for a given applied field of 0.85T.

# Supplementary material for
# Probing pair-breaking mechanisms in proximity-induced hybrid superconducting interfaces

S. Mohapatra, S. Mathimalar, S. Chaudhary, K. V. Raman
*Centre for Interdisciplinary Sciences,*
*Tata Institute of Fundamental Research, Hyderabad 500107, India*

## 1. Calculating differential tunneling conductance

The real solution for the quasiparticle density of states determined above is used to calculate the differential tunneling conductance by taking the convolution of the density of states with the Fermi-Dirac function. Assuming the density of states in the region of NM to be constant, the expression for the differential conductance is given by:

$$G(V) \propto \left.\frac{dI}{dV}\right|_V \propto \frac{d}{dV}\left\{\int_{-\infty}^{\infty} N_S(E)\{f(E) - f(E - eV)\}dE\right\},$$

where V is the applied bias, $N_S$ is the quasiparticle density of states, $f(E,T) = 1/[1 + \exp\{(E - E_f)/k_B T\}]$, being the Fermi-Dirac distribution function at a temperature T and energy E with $k_B$ as the Boltzman constant and $E_f$ as the Fermi level. Equation (6) can be simplified to:

$$G(V) \propto \left.\frac{dI}{dV}\right|_V \propto \left\{\int_{-\infty}^{\infty} N_S(E)\frac{d}{dV}\{f(E) - f(E - eV)\}dE\right\}. \quad (2)$$

The above expression it is used to calculate the differential conductance at any arbitrary value of temperature. The above term reduces to the quasiparticle density of states at very low temperatures.

## 2. Additional figures

### I. Variation in differential conductance to modulations in orbital depairing and spin-scattering interactions

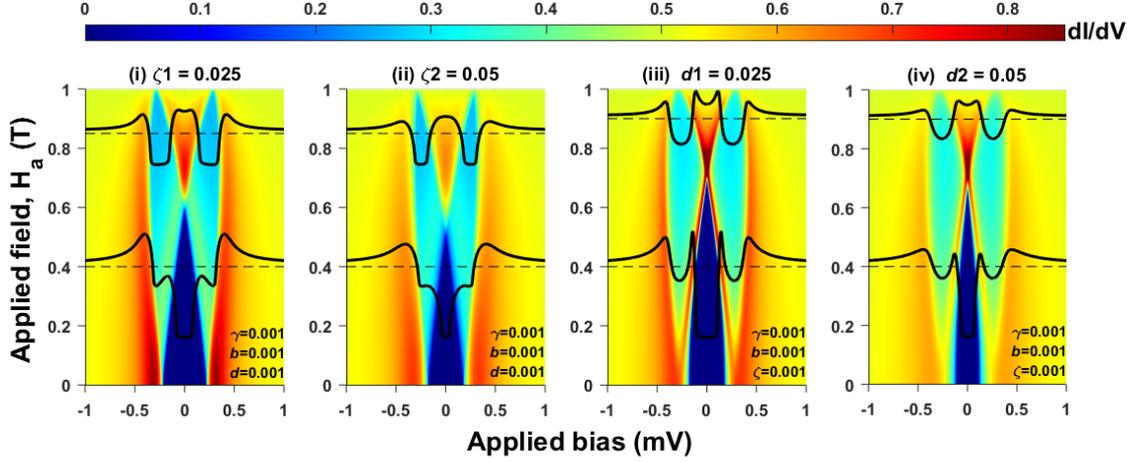

Figure S1**Error! Main Document Only.**: Simulated normalized differential conductance (dI/dV, shown as colorplot in arbitrary units) vs applied bias voltage of TM/SC/I/NM device for varying applied magnetic field with $g_{eff}$ =10 at T=30mK is plotted for increasing depairing contributions of orbital depairing parameter (i & ii) and spin scattering parameter (iii & iv). Superimposed dI/dV line plots (solid line) represent the dI/dV at the field values corresponding to the horizontal line cuts (dashed line).

### II. Observation of hard gap at large spin-orbit strengths

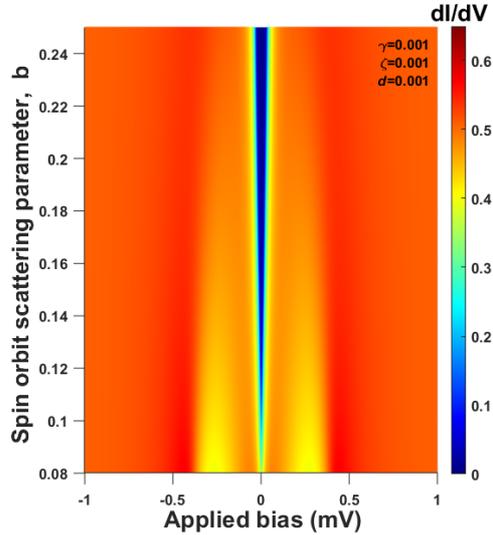

Figure S2: Simulated normalized differential conductance (dI/dV, shown as colorplots in arbitrary units) vs applied bias voltage of an TM/SC/I/NM device with increasing strength of spin-orbit scattering, $b_{so}$, with $g_{eff}$ =10 at T=30mK, taken at 0.85T for a range of $b_{so}$ = 0.08-0.25. The appearance of hard gap at higher values of spin-orbit scattering parameter is evident from the above figure.

## III. Temperature variation showing ZBCP, observed at low temperature, disappearing with the increase in the device temperature

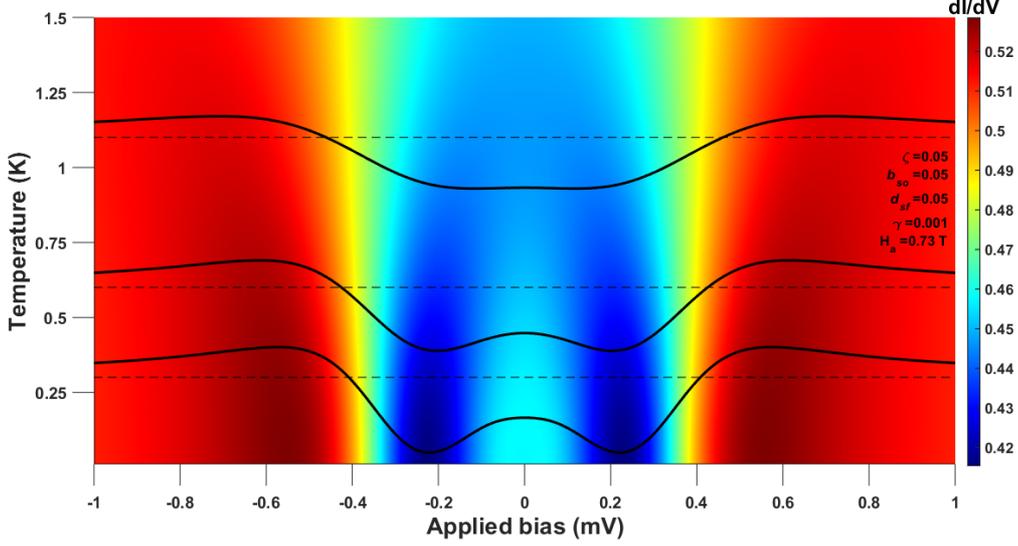

**Figure S3:** Simulated normalized differential conductance (dI/dV, shown as colorplots in arbitrary units) vs applied bias voltage of an TM/SC/I/NM device with increasing temperature is plotted for an applied magnetic field of 730 mT, with $g_{eff}$ =10. Superimposed dI/dV line plots (solid line) represent the dI/dV at the temperature values corresponding to the horizontal line cuts (dashed line). The suppression of the ZBCP at increased temperature ranges is clearly visible.

## IV. Modeling co-existence of proximity induced superconductivity and ferromagnetism by considering an interface exchange field causing Zeeman splitting in zero applied magnetic field.

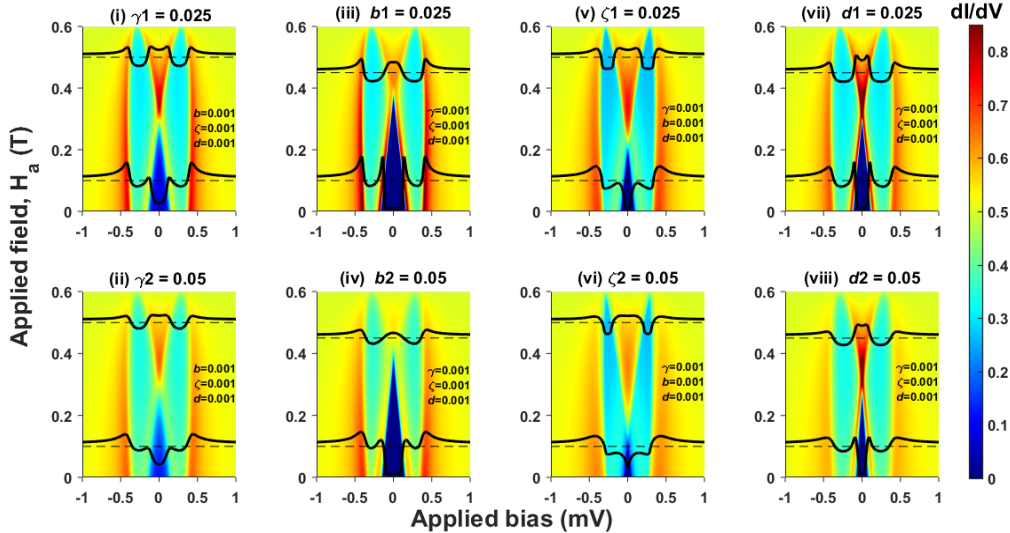

**Figure S4**: Simulated normalized differential conductance (dI/dV, shown as colorplot in arbitrary units) vs applied bias voltage of TM/SC/I/NM device for varying applied magnetic field with $g_{eff}$ =10 at T=30mK and at an internal magnetic field, $H_i$ =0.4 T, is plotted for increasing depairing contributions from Dyne's parameter, $\Upsilon$ (i & ii), spin-orbit scattering parameter, $b_{so}$ (iii & iv), orbital depairing parameter, $\zeta$ (iii & iv) and spin scattering parameter, $d_{sf}$ (vii & viii). Superimposed dI/dV line plots (solid line) represent the dI/dV at the field values corresponding to the horizontal line cuts (dashed line). For each of the subfigures, $\Delta(0)$ = 0.3 meV. Evidently, the formation of the ZBCP and its splitting take place at relatively

lower applied magnetic field, more significantly for smaller depairing parameters, due to the presence of an internal magnetic field. The Zeeman splitting in the absence of applied field is clear from the contrast in the colorplots.